# A High Extinction Ratio Plasmonic Waveguide Polarizer

**Peyman Malekzadeh[1,2], Gholam-Mohammad Parsanasab[1,*], Hamed Nikbakht[1,2], and Ezeddin Mohajerani[2]**

[1]Integrated Photonics Laboratory, Faculty of Electrical Engineering, Shahid Beheshti University, Tehran, Iran

[2]Laser and Plasma Research Institute, Shahid Beheshti University, Tehran, Iran

*gm_parsanasab@sbu.ac.ir

## ABSTRACT

A novel method for increasing the extinction ratio of plasmonic waveguide polarizer has been investigated in this work. Aluminum was coated in a 20 microns multi-strips pattern on the top of a SU-8 waveguide to excite the surface plasmons. Simulation and experimental results have been shown the extinction ratio of the multi-strip polarizer is more than the single strip ones with the same total aluminum-coated length due to the coupling loss between Al coated and uncoated region and also the scattering at the edge of metal. This polarizer was characterized over a wide range of wavelengths and showed a high extinction ratio of 46 dB at 1550 nm and 26.8 dB at 980 nm wavelengths.

## Introduction

The goal of integrated photonic circuits is to transform a bulky free-space setup into a micron/nano-sized chip. For this purpose, each element in the optical setup should have an integrated miniaturized counterpart with a similar function. A polarizer is one of the most frequently used optical elements and has a key role in many optical devices. It is usually used in optical switches, modulators, quantum optics setups, optical gyroscopes, sensors, etc. Therefore, it is one of the most important elements of optical setups and fabricating it in an integrated form has considerable importance.

Multiple approaches have been employed for the fabrication of integrated polarizers including the use of plasmonic structures[1,2], grating-based waveguides[3-5], birefringence materials[6-8], and the absorption of a polarization mode[9,10]. Optical grating-based polarizers can provide a high extinction ratio (ER); however, their fabrication is difficult and requires nanometer resolution[3-5]. Utilizing birefringent materials in the waveguide structure is another method to fabricate polarizers. However, previous experimental results using this method did not provide a high ER[6,7] except for one which had a high insertion loss (IL)[8]. Due to its polarization-dependent absorption, graphene is another option for fabricating polarizers[9,10]. Although graphene is capable of attaining high ERs, the process of fabricating a polarizer with this material is rather complicated. Plasmonic structures have a simple fabrication process and can provide high ERs[1,2]. In these structures, the transverse magnetic (TM) modes can excite the plasmonic mode in the metal and impose losses to it. Since the interaction of the transverse electric (TE) mode with a metal layer is not lossless, these polarizers should be designed in a way to minimize the interaction of this mode with the metal layer. To do this, a dielectric buffer layer is usually placed between the waveguide and the metal layer to minimize the loss in the TE mode. In these polarizers, the length of the metal coating and the thickness of the buffer layer are the key parameters in determining ER[11-13]. In this paper, a novel method is proposed for enhancing ER in plasmonic-based integrated polarizers. In this method, the metal layer is coated in multiple strips on top of the waveguide. Due to the ohmic loss in the metal layer, the coupling loss of the modes, and the scattering at the edges of metal (called 'plasmonic hot spots'), the ER of this polarizer is increased compared to when a solid piece of metal is used. The performed simulations and experiments confirm that using multiple strips leads to more ER than a single piece of metal with the same accumulated length.

Besides, polarizers with different metal strip lengths and buffer layer thicknesses are tested to optimize their performance.

## The working principles

A dielectric buffer layer between the waveguide core and the metal is required to excite surface plasmon polaritons[14,15]. The TM modes excite these oscillations and are hereby absorbed through the ohmic losses in the metal. Besides, the coupling losses between the modes at the beginning and at the end of the aluminum strips and the scattering from the plasmonic hot spots at the edges of the metal reduce the power of the TM mode, while the TE mode has a minor interaction with the metal. To accomplish these conditions, we coated an $MgF_2$ buffer layer on top of a ridge SU-8 waveguide. Then 200-nm-thick aluminum strips with different lengths were patterned on top of the buffer layer, as shown in Fig. 1a.

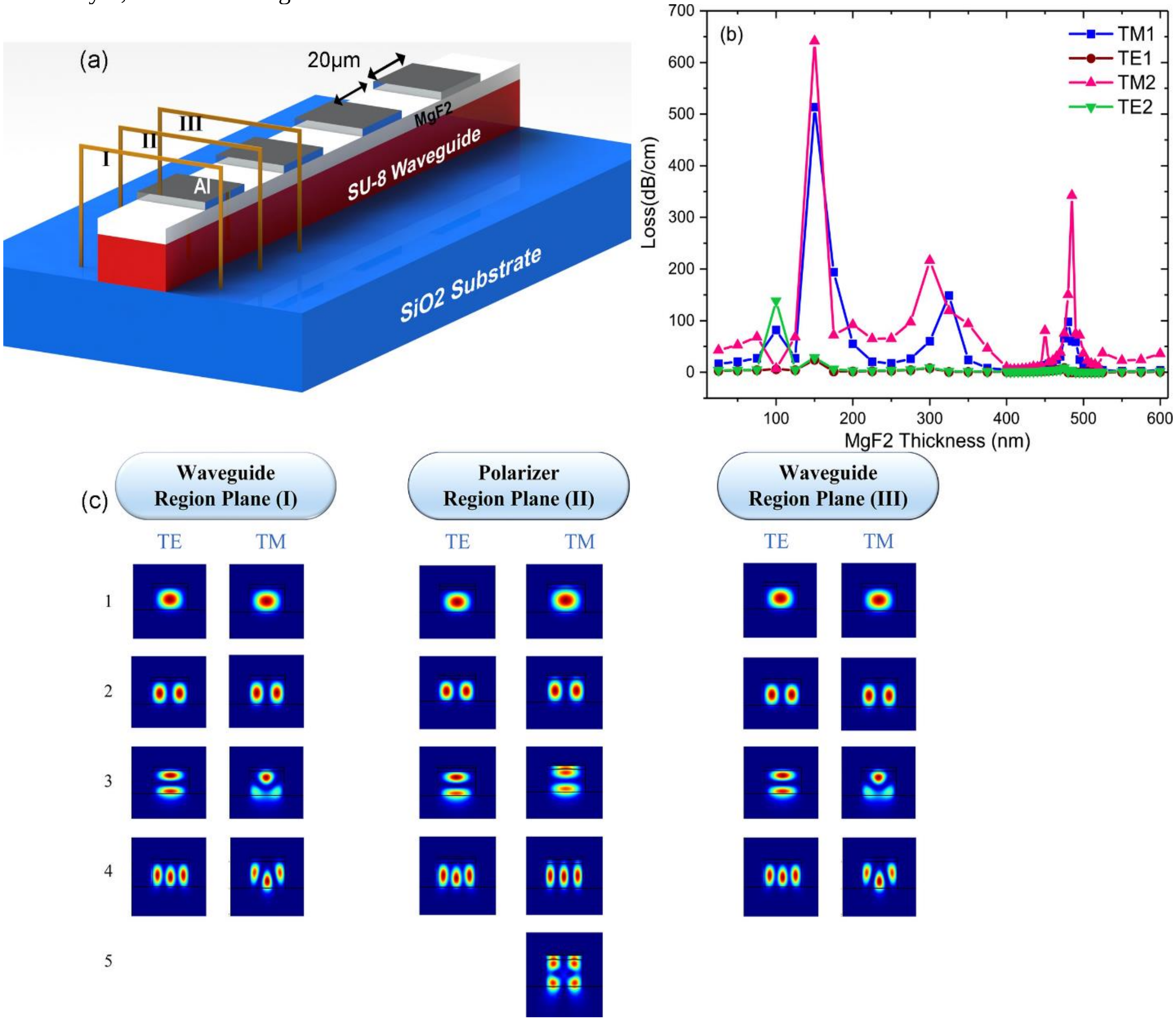

**Figure 1**. (a) The schematic structure of the polarizer. (b) The mode profiles in the Al-coated and uncoated regions.

## Simulation

The first step of the simulation procedure was performed by processing the results from the Mode Solution of the Lumerical software package. Fig. 1b illustrates the TE and TM modes in the regions around an aluminum strip (at planes I, II, and III). The modes before and after the aluminum layer are similar, whereas the modes inside the region with the metal strips are different.
To study the beneficial role of multiple strips, the effect of the coupling coefficients at the entrance and exit of this region are calculated through a simulation. Then a transfer matrix describing the influence of each strip is calculated. Each element of this matrix is calculated by summing the influence of each mode inside the polarizer to transfer the $i^{th}$ input mode to the $j^{th}$ output mode.

$$[E_{ij}]_{m\times n} = \sum_{l=1}^{n} C_{il} e^{-\alpha_l d} e^{-\beta_l d} C_{lj}$$

Where $C_{il}$ and $C_{li}$ are the coupling coefficients from the $i^{th}$ mode of the region without Al coating to the $l^{th}$ mode of the region with Al coating and vice versa, respectively. α is the waveguide attenuation per unit length and β is the propagation constant in the region with Al coating. d is the length of the region with Al coating. m and n are the number of the modes in the regions with and without Al coating, respectively. $E_{ij}$ is the electric field transferred from the $i^{th}$ input mode at the beginning of Al coating to the $j^{th}$ output mode at the end of this region.

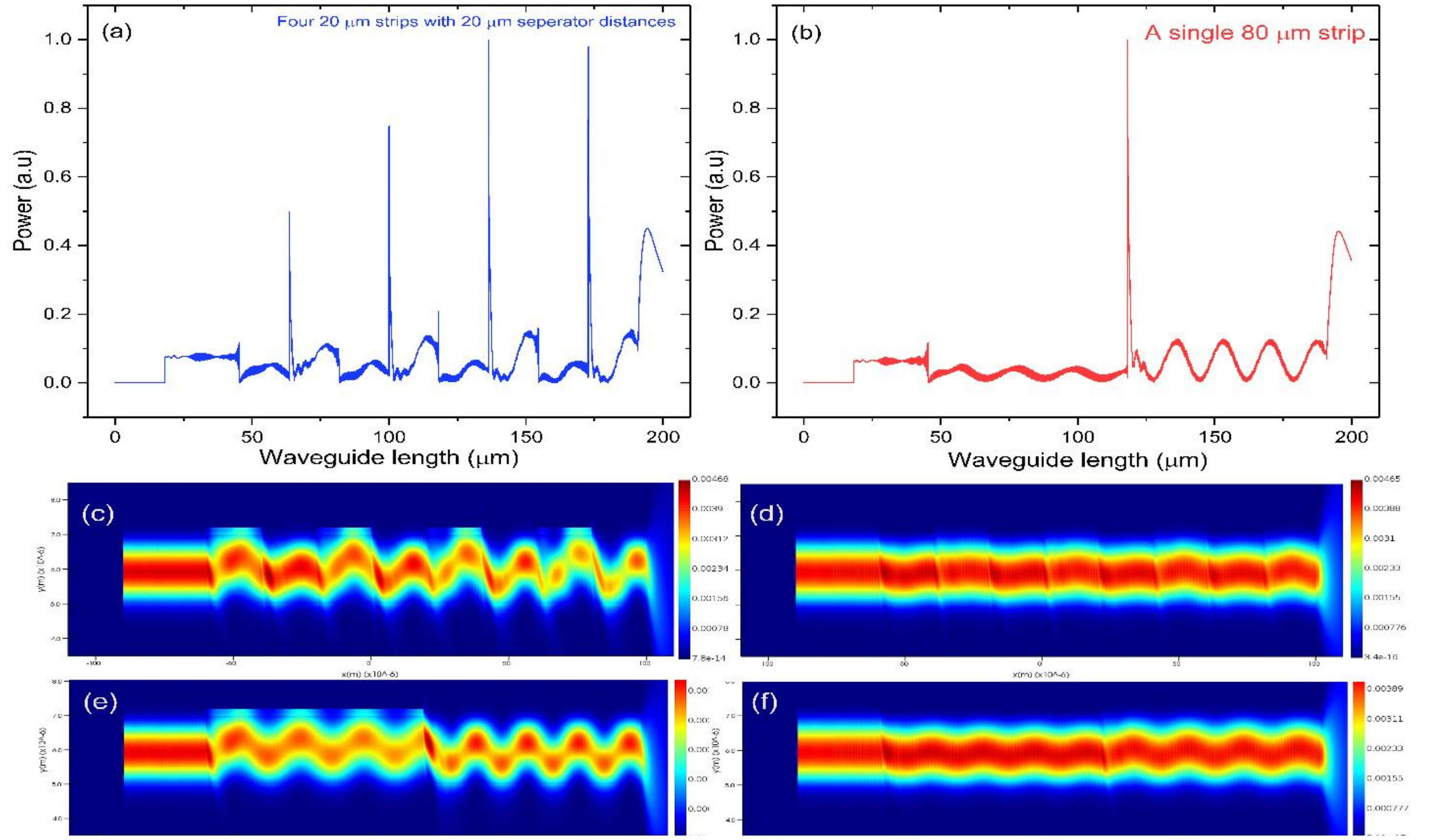

**Figure 2.** Power distribution in the boundary between buffer ($MgF_2$) and metal (Aluminum) layer for (a). four separated 20μm Aluminum strips (b). a single 80 μm strip. To quantitatively compare the difference between the structure with one strip and the structure with four strips, a two-dimensional finite difference time domain (2D-FDTD) simulation was performed at 1550 nm wavelength and the power flux from successive planes perpendicular to the propagation direction was calculated. The results of this simulation have been presented for 20 μm strips with (c). TM and (d). TE mode and for single 80 μm strip with (e). TM and (f). TE mode.

from the i$^{th}$ input mode at the beginning of Al coating to the j$^{th}$ output mode at the end of this region. These calculations confirm more attenuation in the TM modes when the polarizer consists of four aluminum strips with a width of 20 µm compared to one strip with a width of 80 µm. The transmitted power in the structure with four strips is 25% less than that of the structure with a single strip. The thickness of the $MgF_2$ layer was also optimized by comparing the attenuations per unit length calculated in the mode solutions analysis (Fig. 1b). This graph shows that the attenuation of the TM mode is maximized at about 150 nm and 475 nm of $MgF_2$ thickness. However, the attenuation of the TE mode is considerable for the $MgF_2$ layer with a thickness of 150 nm. Therefore, the thickness of 475 nm was chosen for this layer.

FDTD simulation of the structure described is shown in Fig. 2. Plasmonic coupling only occurs for the TM modes. The Sharp peaks of power diagrams are the effect of plasmonic hot spots at the edges of the Al strip (Fig. 2a,b). The simulation of the TM (Fig. 2c) and TE (Fig. 2d) modes in a waveguide with a 200 nm MgF2 buffer layer and four 20 µm Al strips was done. The simulation of the TM (Fig. 2e) and TE (Fig. 2f) modes in a waveguide with a 200 nm MgF2 buffer layer and a 80 µm Al strip was performed. More detailed results are presented in supplementary (Fig. 1).

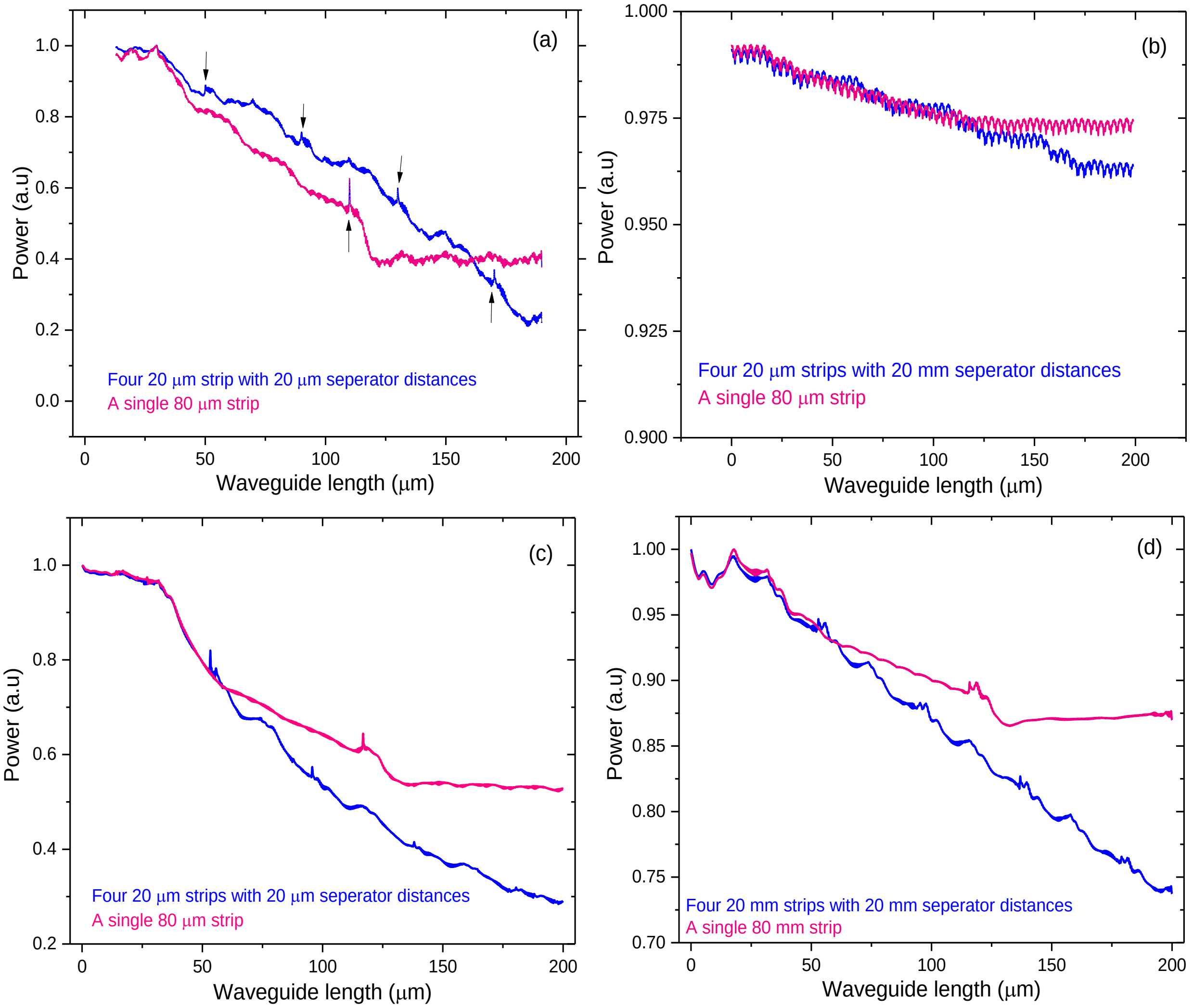

**Figure 3.** Comparing the attenuation of the TM mode in a structure with a 80 µm Al strip (pink) with the four separated 20 µm Al strips (blue) for (a) 200 nm and (b) 500 nm $MgF_2$ layer with fundamental TM mode source and (c) 200 nm and (d) 500 nm $MgF_2$ layer with second TM mode by FDTD simulation at the wavelength of 1550 nm. It can be seen that attenuations of the separated strips are higher than that of the solid strip in all four conditions. The arrows in (a) point to the edge effect that has been discussed before.

Power attenuation (per length) of four-separated strips and a single 80 µm strip was compared in Fig. 3a-d. The first two TM modes are shown for 200 and 500 nm MgF2 layer cases at 1550 nm wavelength. The attenuation of four-separated strips is higher in all cases. Arrows mark the plasmonic hotspot edge effect. By adding Al strips this edge effect increases and it causes more loss for TM mode and higher power attenuation.

## The Fabrication method

An SU-8 photoresist was spin-coated with the speed of 5000 RPM on previously cleaned glass substrates to make a 2 µm thick layer. Afterward, a straight line of this photoresist was exposed with a 405 nm laser beam. After development, a ridge SU-8 waveguide with 4 µm width remained on the substrate[16]. The fabricated waveguides were investigated with an SEM micrograph and then an MgF2 layer with the thicknesses of 200 and 500 nm was coated on the entire structure by thermal evaporation to make the buffer layer between the waveguide and the plasmonic metal layer. Then, the AZ photoresist was spin-coated on the structure and the different strips perpendicular to the waveguide with varying thicknesses for different samples were exposed and developed. After that, an aluminum layer with the thickness of 200 nm was deposited on top of the structure through thermal evaporation and the excess aluminum was removed through the lift-off process in acetone in a sonication bath. Different configurations of Al and $MgF_2$ coatings were fabricated and characterized.

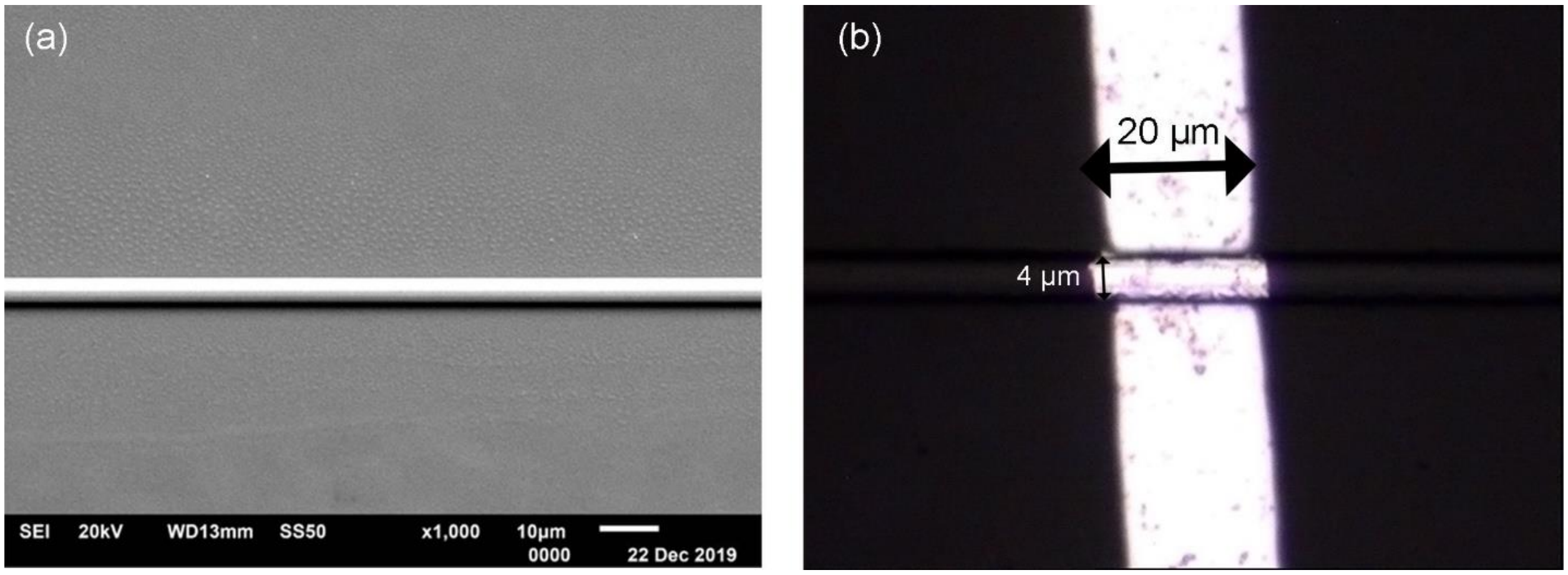

**Figure 4.** (a) The SEM micrograph of the fabricated waveguide with a width of 4 µm and a height of 2 µm. (b) the optical microscope image of the fabricated polarizer after the deposition of a 20 µm long Al strip.

The substrate and the waveguide were broken a few millimeters away from the edges on the two ends to remove deformation at the two ends of the waveguide caused by the edge bead effect. By following this procedure, a flawless waveguide with flat ends can be fabricated which does not require further polishing. Finally, the waveguides were characterized by the coupling of 980 nm and 1525 to 1565 nm tunable laser lights into them.

## The Characterization procedure

In the first step, a laser source with 1550 nm wavelength was end-butt coupled into the waveguide with an optical fiber. The output of the waveguide was monitored by an infrared camera. The change in the orientation of the polarizer showed some variations in the transmitted light from the polarizer. The minimum intensity was observed when the free-space polarizer was aligned parallel to the substrate surface (Fig. 5a). This test confirmed that the light exiting from the waveguide was TE-polarized.

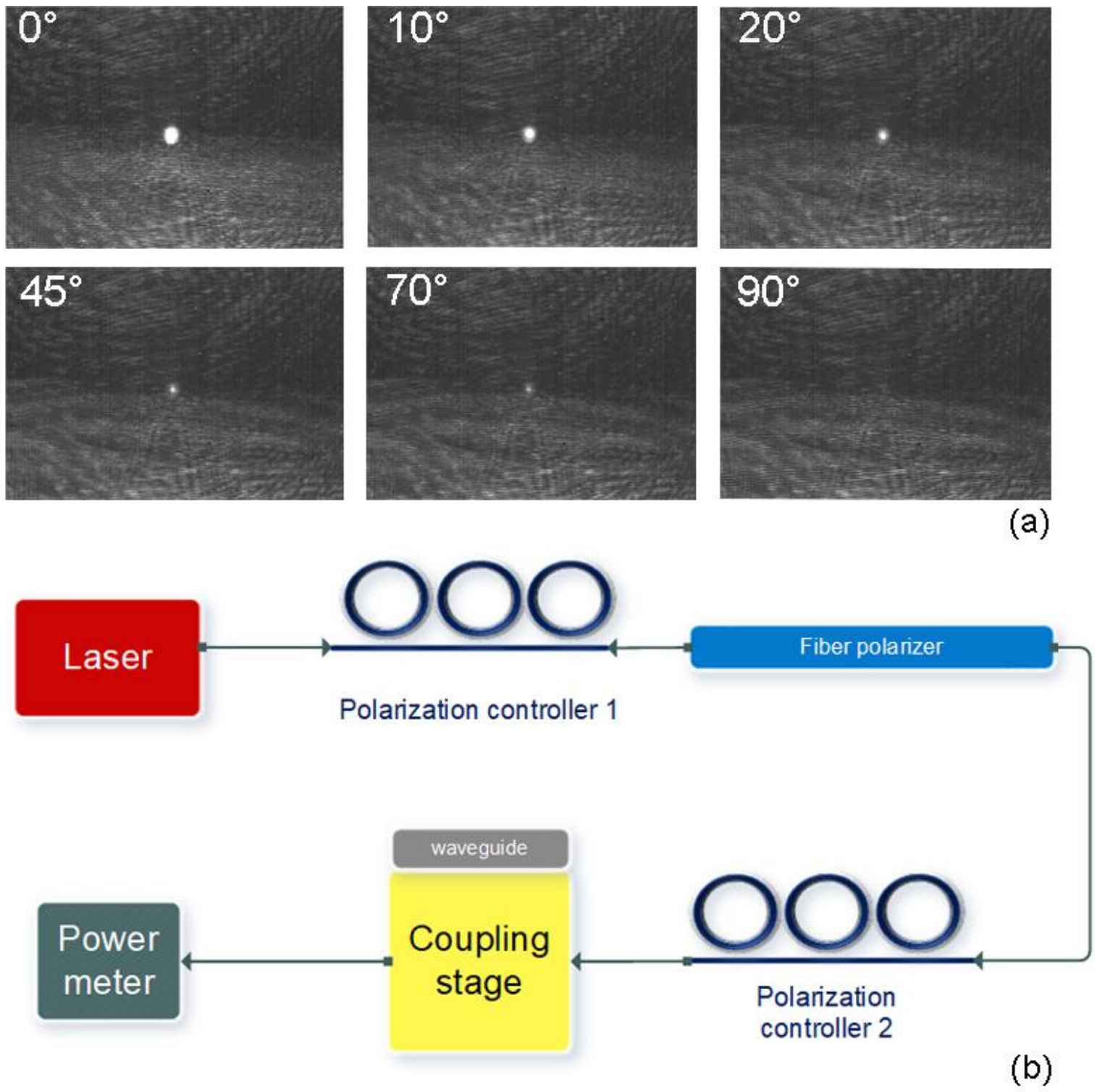

**Figure 5.** (a) polarizer angles. (b) The optical characterization setup of the fabricated polarizer.

In the next step, the fabricated polarizer was characterized using a polarization measurement setup. In (Fig. 5b). Light from a tunable laser (Thorlabs, TLX1), which was partially polarized, was passed through a fiber polarization controller. Afterward, the extinction ratio of the light was increased by passing it through a commercial fiber polarizer and the polarization controller was adjusted to maximize the transmitted power. The high-ER light exiting from the commercial polarizer was passed through another fiber polarization controller and was then coupled to the waveguide using end-butt coupling from a cleaved fiber to the waveguide. The light was coupled to another cleaved optical fiber at the exit of the waveguide and this power was measured with a power meter (EXFO FOT-930). The output power of the waveguide was maximized and minimized by adjusting the second polarization controller to measure the extinction ratio of the fabricated polarizer. The maximum and minimum powers through which the extinction ratio can be calculated were recorded. This procedure was repeated for different wavelengths by changing the wavelength transmitted from the tunable laser and the maximum, minimum, and extinction ratio were determined over a wide spectral range. the pictures of this setup are provided in supplementary fig 2.

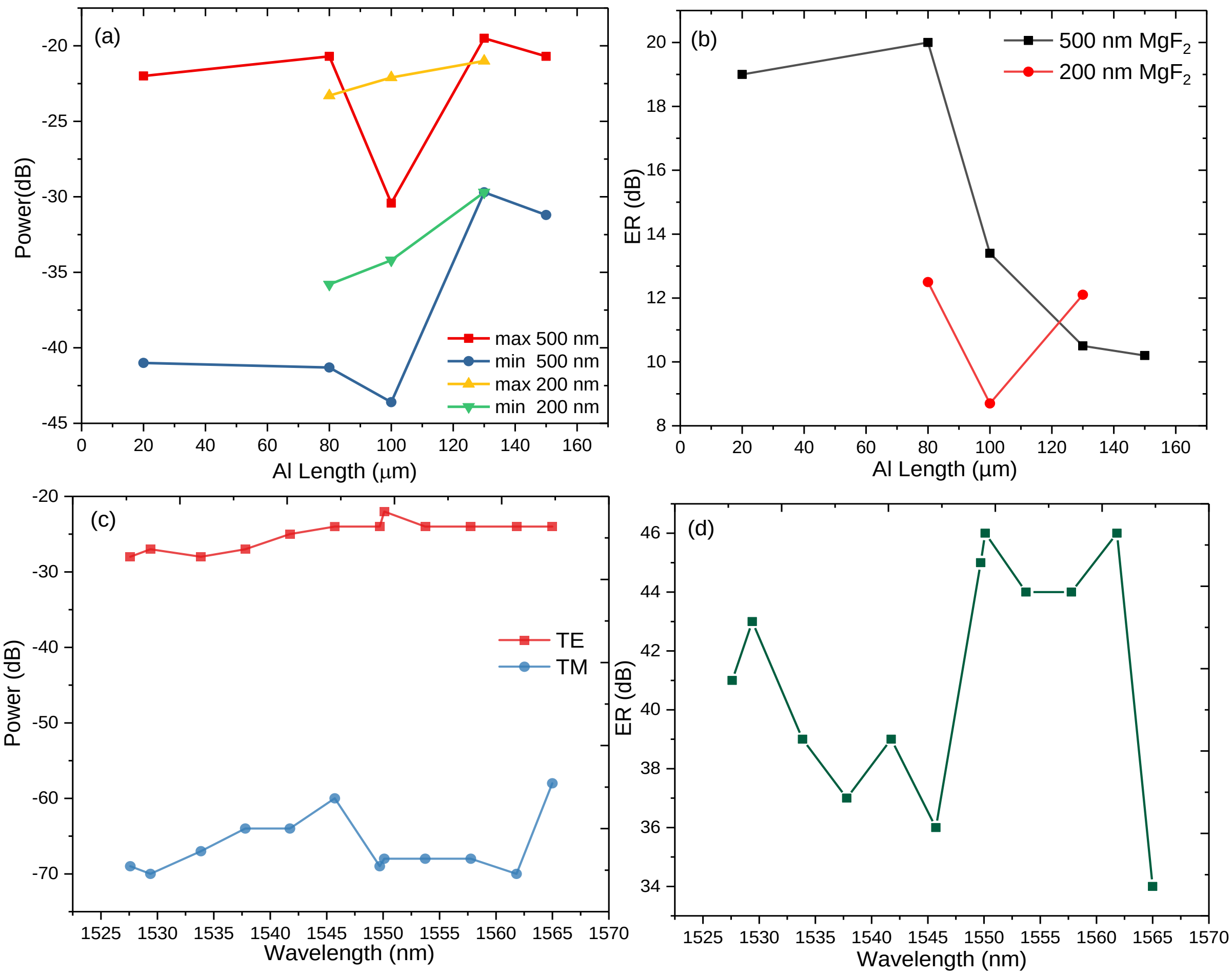

**Figure 6.** (a) The maximum and minimum powers from the polarizers versus the width of aluminum strips for different thicknesses of the buffer layer (b) The ER of the transmitted power in (a). (c) The maximum and minimum transmitted powers from the polarizer consisting of four strips of aluminum with a width of 20 µm. (d) The ER of the transmitted power in(c). the measurement error for all measurements is about 0.5 dB.

## Results and discussion

The first result was acquired by adding one strip of aluminum with different lengths to the top of the waveguide. This experiment was performed for the buffer layers with the thicknesses of 200 and 500 nm in Fig 6. As can be observed from this (Fig. 6b), the ER decreased for aluminum length larger than 80 µm. Besides, the overall ER for the buffer layer with a thickness of 200 nm was lower than the one with a thickness of 500 nm.

In the next experiment, the number of strips was increased to four and the ER measurements were repeated. Fig. 6c shows the maximum and minimum transmitted powers from this polarizer following the characterization procedure described above. The maximum power varied from -22 to -27 dB. This variation was caused by the variation of the power transmitted from the source and the changes in the coupling coefficients for different wavelengths. The minimum power varied from -70 to -58 dB. Fig. 6d shows the extinction ratios calculated from the maximum and minimum powers. The maximum ER of 46 dB is observed in this graph at 1550 nm. The characterization procedure was repeated for the wavelength of 980 nm and the ER value of 26.8 dB was obtained. In addition, the ER value for the uncoated waveguides was obtained as 0.6 dB.

| Year | WG Material | Wavelength | Mechanism | ER | References |
|---|---|---|---|---|---|
| 2019 | **SU-8** | **1510 nm** | **Birefringence** | **14 dB** | **6** |
| 2012 | **Silicon** | **1550 nm** | **Plasmonic** | **15 dB** | **12** |
| 2010 | **GaAs-InP** | **1550 nm** | **Photonic crystal** | **16 dB** | **17** |
| 2012 | **Resin polymer** | **1310 nm** | **Absorbance** | **19 dB** | **9** |
| 1997 | **Polymer** | **1550 nm** | **Birefringence** | **21 dB** | **7** |
| 2008 | **Silicon** | **1315 nm** | **Plasmonic** | **30 dB** | **2** |
| 2019 | **Germanium** | **1540 nm** | **Grating** | **30 dB** | **4** |
| 2006 | **LiNbO3** | **1550 nm** | **Plasmonic** | **32 dB** | **18** |
| 1998 | **Silica** | **1310 nm** | **Birefringence** | **39 dB** | **8** |
| 2014 | **Silicon** | **1550 nm** | **Grating** | **40 dB** | **5** |
| 1992 | **K+-Na+** | **836 nm** | **Plasmonic** | **43 dB** | **1** |
| Our work | **SU-8** | **1550 nm** | **Plasmonic** | **46 dB** | **----** |
| 1999 | **PMMA** | **820 nm** | **Metal grating** | **50 dB** | **3** |

**Table 1**. A comparison of the fabricated integrated polarizers in the recent years

## Conclusion

In this paper, an integrated polarizer with a high ER was proposed and fabricated with a simple lithography procedure. It was shown that increasing the number of metal strips on top of the waveguide with the same cumulative length increased the ER of the polarizer. The fabricated polarizer works in a wide range of wavelengths and is capable of creating ERs as high as 46 dB. This is the highest value at the wavelength of 1550 nm reported in the literature for integrated polarizers. This is easy to fabricate and use method to apply in SU-8 integrated photonic devices and other platforms.

## Competing interests

The authors declare no competing interests.